\documentclass[12pt]{elsart}
\usepackage{graphicx}
\usepackage{hyperref}
\usepackage{setspace}
\usepackage{longtable}

\begin{document}
\begin{frontmatter}
\title{Velocity plateaus and jumps in carbon nanotube sliding}
\author{Xiao-Hua Zhang$^{a}$},
\author{Ugo Tartaglino$^{b}$\corauthref{cor}},
\ead{tartagli@sissa.it} \corauth[cor]{Corresponding Author.}
\author{Giuseppe E. Santoro$^{b,c}$, and}
\author{Erio Tosatti$^{b,c}$}
\address{$^a$ Surface Physics Laboratory and Department of Physics,
Fudan University, Shanghai 200433, China}
\address{$^b$ International School for Advanced Studies (SISSA),
and INFM-CNR Democritos National Simulation Center, Via Beirut 2-4,
I-34014 Trieste, Italy}
\address{$^c$ International Centre for
Theoretical Physics (ICTP), P.O.Box 586, I-34014 Trieste, Italy}

\maketitle
\begin{abstract}

The friction between concentric carbon nanotubes sliding one inside 
the other has been widely studied and simulated, but not so far using
external force as the driving variable. Our molecular dynamics (MD) 
simulations show that as the pulling force grows, the sliding velocity 
increases by jumps and plateaus rather than continuously as expected. 
Dramatic friction peaks (similar to that recently noted by Tangney {\it et al.} in Phys. Rev. Lett. 97 (2006) 195901)
which develop around some preferential sliding velocities, are at 
the origin of this phenomenon. The (stable) rising edge of the peak 
produces a velocity plateau; the (unstable) dropping edge produces a 
jump to the nearest stable branch.
The outcome is reminiscent of conduction in ionized gases, the plateau corresponding to a current 
stabilization against voltage variations, the jump corresponding to a discharge or breakdown.
\end{abstract}

\end{frontmatter}

\section{Introduction}

Several years ago Zettl and others showed that nanotubes can be made to 
slide one inside another \cite{Zettl_SCI00,Zettl_NAT03,Zettl_PRL06,Bachtold}. 
The sliding friction proved to be rather low, and the inner tube could even 
execute weakly damped oscillations out and in the outer one.
 
Assuming nanotube sliding friction to imply no wear, the remaining sources
of friction are through {\it i)} dissipation of phonons from and through the 
outer tube to the substrate onto which it is deposited, and {\it ii)} electronic 
friction processes, corresponding to any extra internal resistivity arising inside 
each tube due to scattering caused by the presence of the other tube \cite{Persson_book}. 

Little is known about nanotube electronic friction, and only indirect
evidence of its existence has been identified so far \cite{Persson_PRB04}.
Phononic nanotube friction, which on the other hand is the focus of our 
attention here, has been theoretically studied by various authors,
particularly using classical molecular dynamics (MD) simulations 
\cite{Servantie_PRB06,Guo_PRB,Servantie_PRL06,Tangney_PRL04}. 

The main source of tube-tube phononic friction was so far believed 
to be the inner nanotube terminations \cite{Tangney_PRL04},
while the ``bulk'' friction, arising from mutual sliding of interior tube walls, is often 
considered to be irrelevant. Because of that, many simulation studies did not attempt to 
describe realistically the wall-wall interactions, sometimes assumed 
of a crude Lennard-Jones (LJ) type, with parameters which may underestimate
the tube-tube corrugation and thus the bulk friction.

Recently Guo \cite{Guo_PRB} remedied that aspect and
focused attention on bulk nanotube friction, adopting more accurate
interactions, such as those parametrized on ab-initio energies by
Kolmogorov and Crespi \cite{KC2005}.    

In a recent exciting development, Tangney {\it et al.} \cite{Tangney_PRL06}
showed that tube-tube sliding friction can grow very large for some special sliding
velocity, for example $v\approx 1100 m/s$. This finding was rationalized by
invoking a resonance condition between the ``washboard frequency'' 
$\omega_w = 2 \pi v/a$ (where $a$ is the inter-tube potential corrugation period) 
and particular phonon frequencies of the outer tube (radial breathing and longitudinal acoustic)
which happen to possess a group velocity approximately equal to the sliding velocity $v$. 

Methodologically, it would moreover be more interesting to describe nanotube friction 
as a function of the applied sliding {\em force} $F$, rather than of
some externally imposed velocity $v$. Force is in fact the external control 
parameter in friction, whereas sliding velocity is rather an outcome than an input. 
Thus for example, phenomena like stick slip occur in the real situation 
of an externally applied force, but they would not for a hypothetical externally 
applied velocity. 

We describe here the simulated behavior of nanotube (phononic) sliding friction, now 
as a function of applied force $F$, pulling the inner tube against the outer one. 
We use realistic tube-tube interactions in otherwise standard 
classical MD simulations. Because we aim at describing purely bulk friction, 
we slide infinite tubes one inside the other, without any terminations whatsoever.

Preliminary results of our simulations show several novel features. 
The inner tube sliding velocity does not grow smoothly for increasing pulling force. 
There are instead large plateaus where the velocity $v$ stays nearly constant
in spite of a growing force $F$. Plateaus terminate with velocity jumps, typically
to another velocity plateau. A second aspect concerns rotation. Leaving the inner
tube free to rotate, there is no proper rotation in nonchiral tubes, where there may
still be pseudorotations. On the other hand, there is generally a true rotation of
the inner tube in the chiral case. Here we find that the inner tube angular velocity 
also grows by plateaus and jumps, which occur in coincidence with those of the 
linear velocity. 

\begin{figure}
\includegraphics[width=0.80\textwidth]{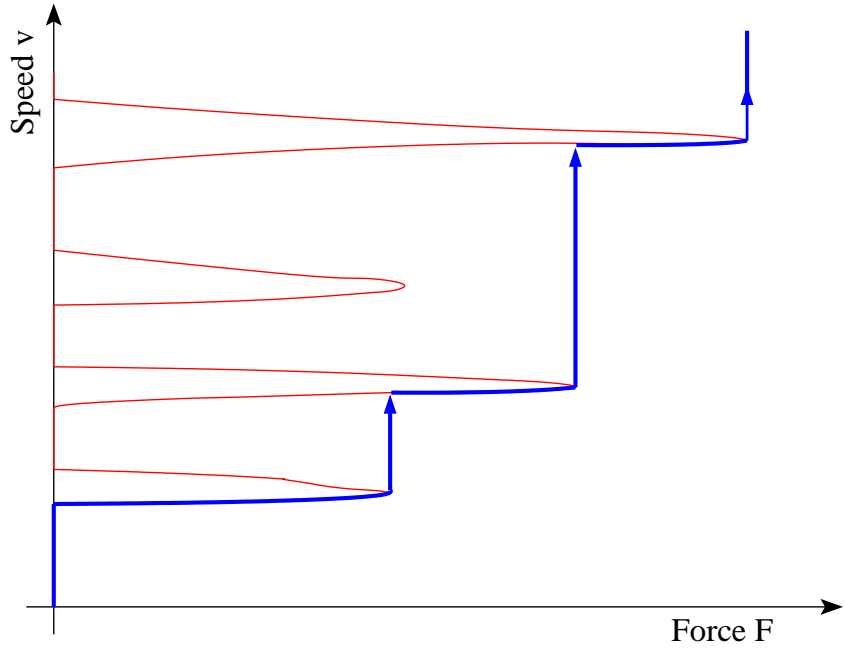} 
\caption{ \label{sketch:fig}
(Color online) Sketch of a $F-v$ characteristics (typically S shaped, i.e., with peaks sitting
on the ordinate axis), and a typical $v$ versus $F$ curve obtained by ramping $F$
(thick solid line), with plateaus in $v$ followed by abrupt jumps (vertical arrows).
}
\end{figure}

The plateaus and jumps are rationalized as the natural outcome of frictional
peaks for growing velocity, the stable (rising) side of the
peak leading to a plateau, the dropping (unstable) side leading to a jump,
as depicted schematically in Fig.~\ref{sketch:fig}.
The presence, in a general tube 
force-velocity characteristics, of several peaks accounts for the multiplicity 
of plateaus and jumps. For each given pair combination of sliding tubes, their 
frictional peaks need to be mapped out. The peaks are ostensibly due to parametric 
excitation of vibrational modes in the double wall nanotube caused by the inner tube sliding,
but the details of these parametric excitation need to be worked out in more detail.

\section{Infinite nanotube sliding: Molecular dynamics simulations methods}

The simulations presented here are restricted to two concentric nanotubes, 
a strongly commensurate case --- where the inner tube is a (5,5) armchair, and the outer 
one is (10,10) ---, and a nearly incommensurate case, with a chiral (11,2) nanotube 
sliding inside a (12,12) one, as illustrated in Fig.~\ref{nanotubes:fig}.
The (5,5)--(10,10) simulation cell comprises 160 (inner) + 320 (outer) carbon atoms.  
The (11,2)--(12,12) simulation cell comprises 196 (inner) + 336 (outer) carbon atoms.
The two concentric nanotubes have a common $z$-axis, but have no terminations: 
they are connected across the boundary of the simulation cell through periodic 
boundary conditions (PBC) in the $z$-direction.
This implies that we concentrate on bulk sliding, which is naturally the 
relevant source of the frictional peaks \cite{Tangney_PRL06}.

The inner tube is pulled along the axis with a force $F$ acting on each atom, 
while the outer tube is subject to two constraints:
one that fixes the position of its center of mass, and
another that prevents the rotation around the $z$-axis. 
(These are global constraints: every single atom of the nanotubes is free to move.)
A Berendsen thermostat \cite{Berendsen} acts on the atoms of the outer tube alone, 
maintaining it at a temperature $T$, thus removing the frictional energy. 
This is intended to mimic the behaviour of the real system, where the inner tube 
dissipates only through its interaction with the outer tube, in turn thermalized by direct contact 
with the substrate.

The potential used in modeling intra-tube forces is a standard Tersoff-Brenner \cite{Brenner},
while inter-tube forces are described by the most recent Kolmogorov-Crespi potential \cite{KC2005}. 
This choice was made following Ref.~\cite{Guo_PRB}, who argued that a LJ interaction would underestimate
the corrugation of the inter-tube potential, as is known to be the case in graphite \cite{Guo_PRB}. 

\section{Results}

\begin{figure}
\includegraphics[width=0.8\textwidth]{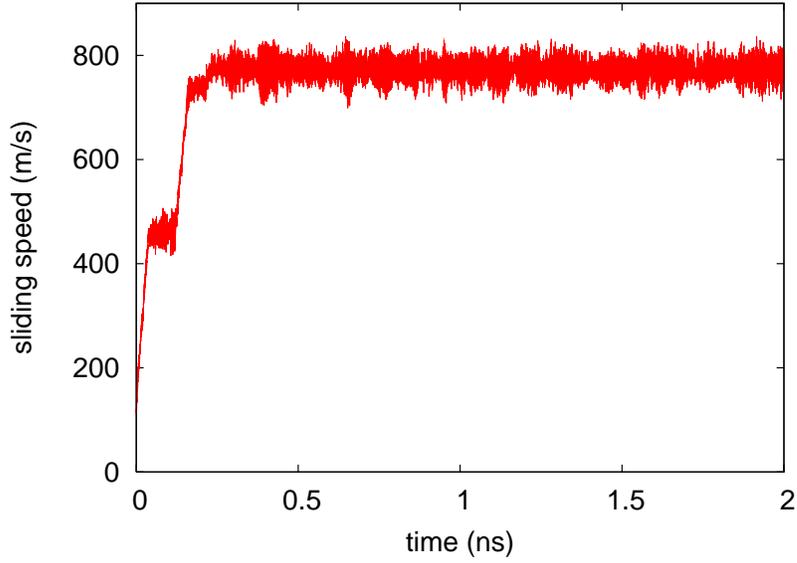} 
\caption{  \label{speedjumps300K}
Speed jumps and plateaus of the inner tube in a commensurate (5,5)-(10,10) case,
when the inner tube is pulled with a constant force $F=0.12$ meV/\AA\ per atom.
Here $T=300$ K.
}
\end{figure}

Fig.~\ref{speedjumps300K} shows the time-dependence of the (5,5) inner-tube velocity
when sliding inside a commensurate (10,10) outer tube under a fixed force $F$. 
The starting point was from rest. The outer tube is maintained at $T=300$ K and 
impeded to rotate or translate globally. Two velocity plateaus 
are seen at $v\approx 450$ m/s and $v\approx 780$ m/s, separated by  
large velocity jumps.

\begin{figure}
\includegraphics[width=0.8\textwidth]{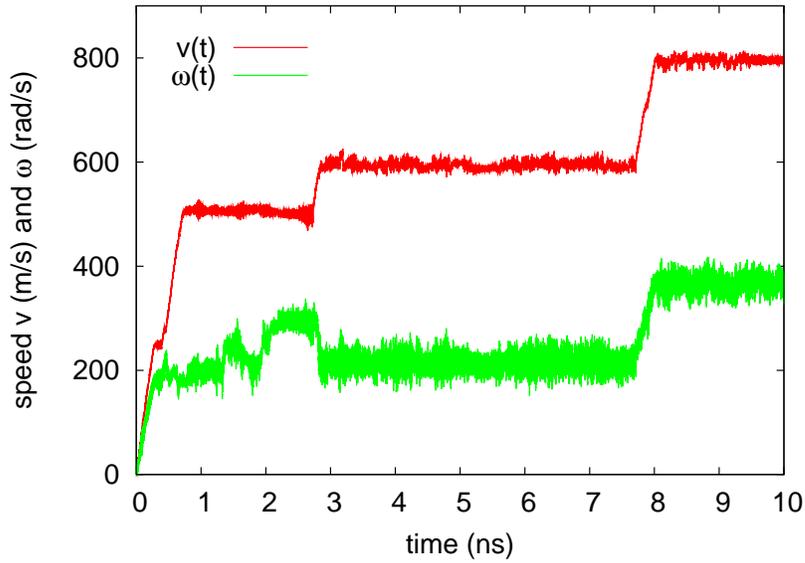} 
\caption{  \label{speedjumps50K}
Chiral nanotube (11,2) inside the (12,12) nanotube pulled with a
constant force $F=0.013$ meV/\AA\ per atom.
The speed increases by jumps. The motion is accompanied
by a rotation induced by the chirality of the inner tube.
}
\end{figure}

A similar occurrence of plateaus and jumps is observed for the 
incommensurate (11,2)--(12,12) case, shown in Fig.~\ref{speedjumps50K} 
(here the outer tube is at $T=50$ K), with the extra ingredient that the 
translational velocity plateaus and jumps are accompanied by a corresponding 
set of rotational plateaus and jumps in the global angular velocity of the inner (chiral) tube. 

We have just two examples here; but all indications suggest
that these remarkable plateau-jump frictional phenomena are ubiquitous 
in nanotubes. In order to understand them we will concentrate here on the simpler 
commensurate (5,5)--(10,10) case, leaving the more complex chiral case for 
future studies. We carried out additional simulations, now at at fixed speed, 
to explore in more details the $v$--$F$ characteristics.
The simulations are performed by imposing a time-dependent force $F(t)$,
equal for all atoms, which enforces the constraint that the inner tube
has to move with a given velocity $v$, following the general procedure of Ref.~\cite{Andersen_JCP83}. 

\begin{figure}
\includegraphics[width=0.8\textwidth]{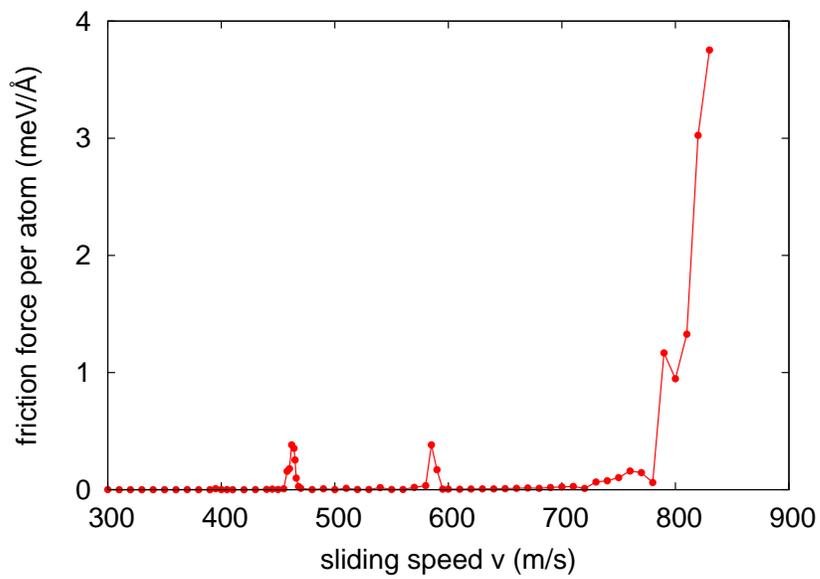} 
\caption{  \label{friction_vs_speed}
Simulations at constrained sliding speed for the commensurate case. 
The friction force is negligible everywhere except around some velocities corresponding
to the speed plateaus observed in the simulations at fixed pulling force.
}
\end{figure}

After an initial transient, the force $F(t)$ effectively balances (on average)
the friction forces due to the outer tube, and provides, therefore, a direct measurement
of $F_{\rm friction}$, $F_{\rm friction}=\overline{F(t)}$, the bar indicating a time-average.
Fig.~\ref{friction_vs_speed} shows the results obtained for $F_{\rm friction}$ versus $v$,
at temperature T=50 K. There is a clear peak of $F_{\rm friction}$ at $v\approx 450$ m/s,
followed by another peak at $v\approx 595$ m/s, and by a large peak at $v\approx 800$ m/s.

Comparison of the frictional peaks of Fig.~\ref{friction_vs_speed} with the two plateaus and jumps
found by force pulling in Fig.~\ref{speedjumps300K} leads to a clear identification. The plateaus are due
to the rising edge of the frictional peak. So long as velocity grows, even marginally,
for growing force, the system is mechanically stable. As soon as the force exceeds
the peak value, however, the velocity should decrease for increasing force, and that causes
a mechanical instability. 
The instability is then resolved by a jump to a higher stable branch.

\section{Discussion}

The main result of this work is the observation that pulling the inner nanotube with a well
defined force of increasing (or decreasing) value generally leads to anomalous
frictional sliding, characterized by plateaus and by jumps of the inner tube velocity.

That behavior can be traced to the presence of peaks in the force-velocity
characteristic. The frictional peaks correspond to resonant excitation of modes 
of the two-wall nanotube, when in the sliding configuration. 
According to Tangney {\it et al.} \cite{Tangney_PRL06}, who recently found one such peak, 
these should be modes of the outer nanotube at the washboard frequency with group velocity $\approx v$
(the sliding velocity). 
Here, however, the washboard frequencies are 3.7 THz and 6.3 THz for the 450 and 780 m/sec plateaus, 
respectively, whereas we observe, by analysing the velocity Fourier transforms (not shown), that the 
outer tube has large resonances at 0.14 THz and 0.65 THz, respectively. Thus the 
simple prescription seems not to work, and we must defer a detailed study of 
this aspect to further work.

The mechanical instability under external force pulling is a novel aspect
here, although actually quite a familiar one in a more general context. 
In electrical engineering, for example, an instability of a similar kind 
actually presents itself in all I-V systems, characterized by S-shaped  
negative differential resistance. A simple analog here would be conduction 
through a gas, where the current remains the same for a large plateau of 
voltages, to jump suddenly up at the onset of current-induced ionization, 
and consequent discharge.  

The phenomena presented in the present paper seem quite ubiquitous, and rather insensitive to
many of the details of the simulations, and are therefore likely to occur in experiments.

\section*{Acknowledgments}

\begin{figure}
\begin{center}
\includegraphics[width=0.4\textwidth]{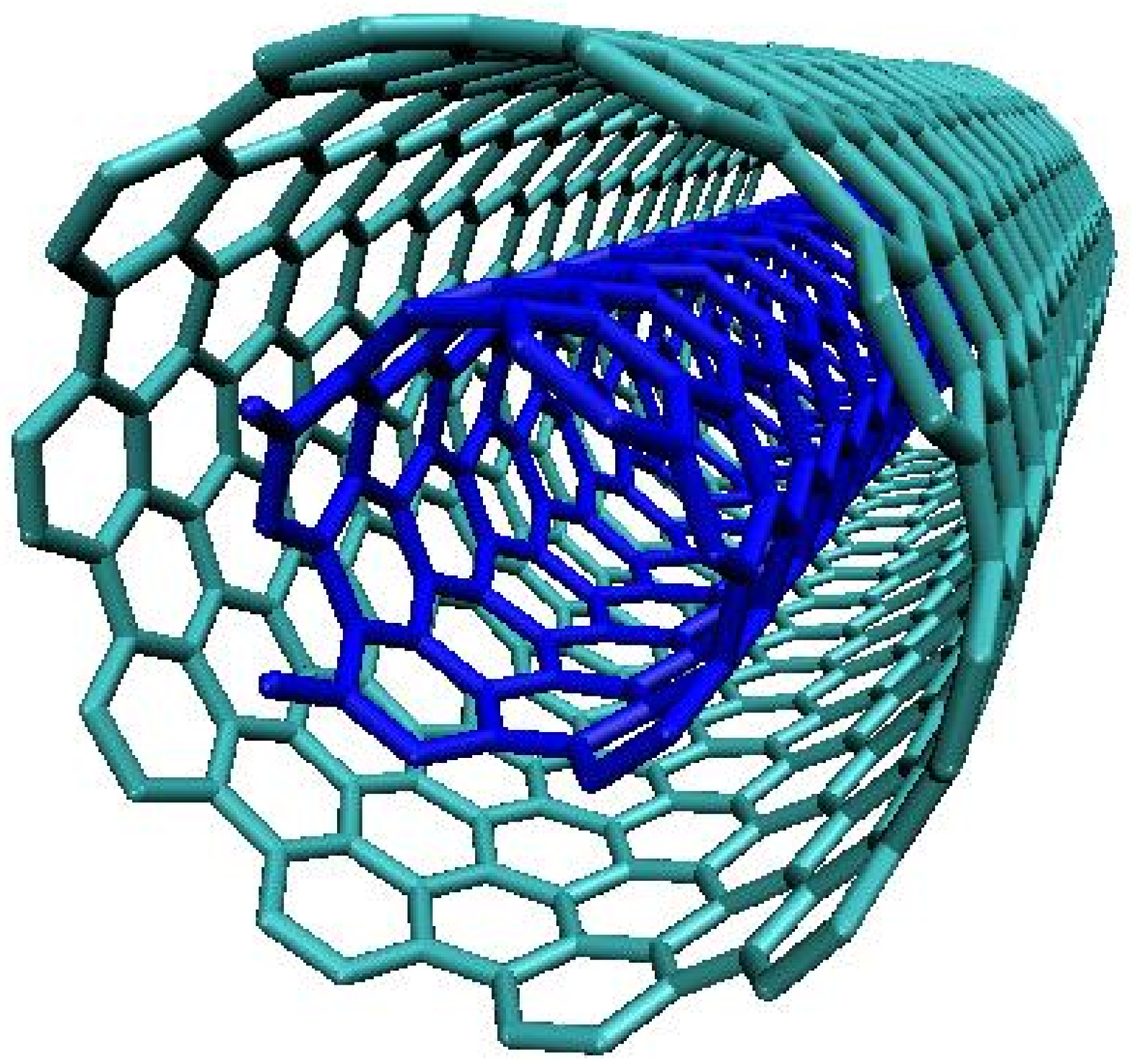} 
\includegraphics[width=0.4\textwidth]{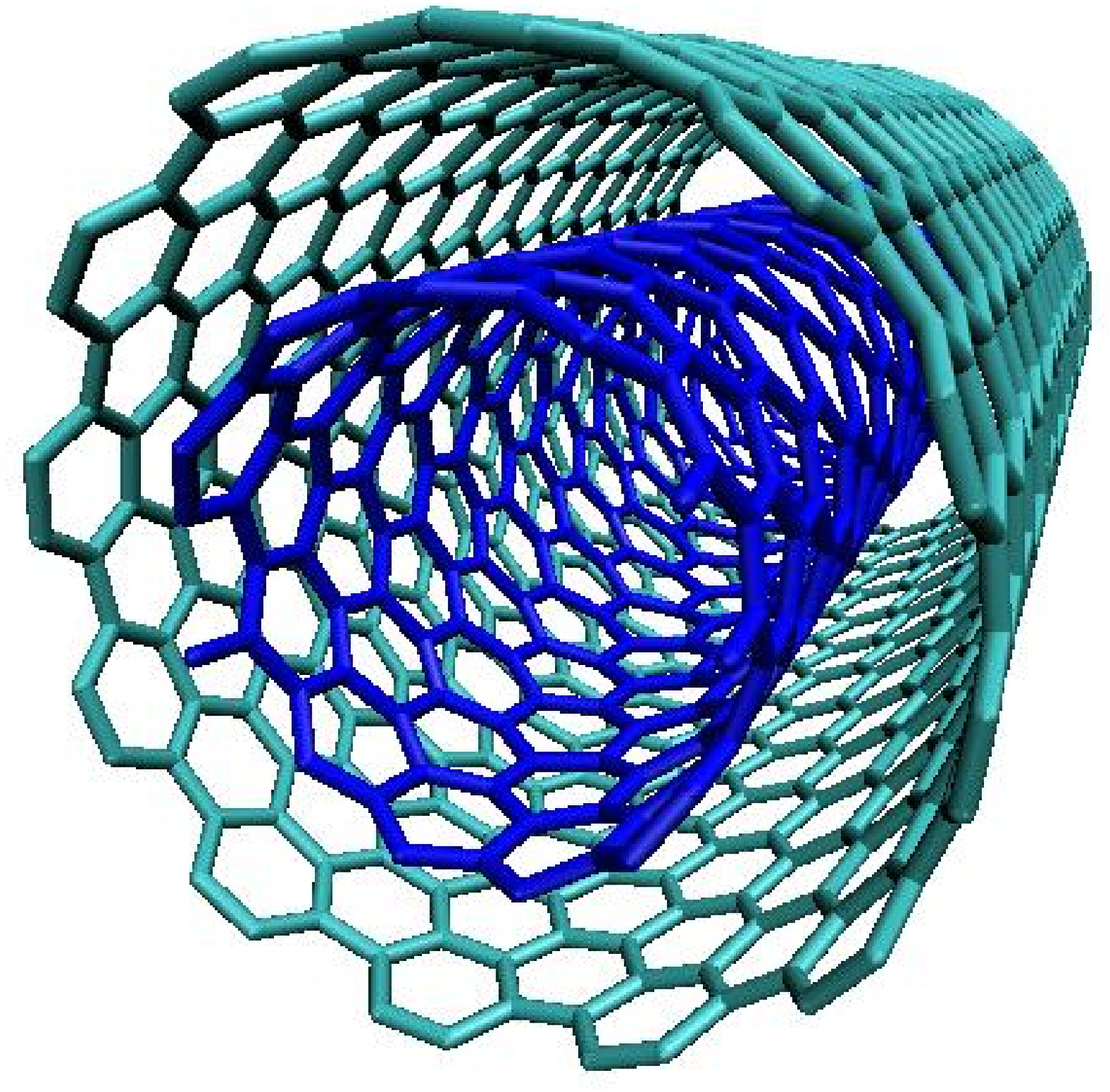} 
\end{center}
%
\caption{\label{nanotubes:fig} (Color online) Left: The commensurate case, a (5,5) nanotube 
inside a (10,10). Right: The incommensurate (chiral) case, a (11,2) nanotube inside a (12,12).}
\end{figure}

One of us (XHZ) is grateful to Professor X.G. Gong for encouragement.
XHZ also acknowledges SISSA for a fellowship enabling him to visit Trieste,
where this work was started. At SISSA/Democritos this research was partially supported by 
MIUR Cofin 2004023199, FIRB RBAU017S8R, and RBAU01LX5H.
Atomic structure images in Fig.~\ref{nanotubes:fig} have been produced with
VMD (http://www.ks.uiuc.edu/Research/vmd/) \cite{HUMP96}.

%


\begin{thebibliography}{10}

\bibitem{Zettl_SCI00}
{J. Cummings and A. Zettl, Science {\bf 289}, 602 (2000)}.

\bibitem{Zettl_NAT03}
{ A.M. Fennimore, T.D. Yuzvinsky, Wei-Qiang Han, M.S. Fuhrer, J. Cumings and A.
  Zettl, Nature {\bf 424}, 408 (2003)}.

\bibitem{Zettl_PRL06}
{ K. Jensen, \c{C}. Girit, W. Mickelson and A. Zettl, Phys.\ Rev.\ Lett.\ {\bf
  96}, 215503 (2006)}.

\bibitem{Bachtold}
{B. Bourlon, D.C. Glattli, C. Miko, L. Forr\'o and A. Bachtold, Nano Lett.\
  {\bf 4}, 709 (2004)}.

\bibitem{Persson_book}
{B.N.J. Persson {\it Sliding Friction: Physical Principles and Applications}
  2nd ed., Springer, Heidelberg (2000)}.

\bibitem{Persson_PRB04}
{B.N.J. Persson, U. Tartaglino, E. Tosatti, and H. Ueba, Phys.\ Rev.\ B {\bf
  69}, 235410 (2004)}.

\bibitem{Servantie_PRB06}
{J. Servantie and P. Gaspard, Phys.\ Rev.\ B {\bf 73}, 125428 (2006)}.

\bibitem{Guo_PRB}
{W. Guo, W. Zhong, Y. Dai and S. Li, Phys.\ Rev.\ B {\bf 72} 075409 (2005)}.

\bibitem{Servantie_PRL06}
{J. Servantie and P. Gaspard, Phys.\ Rev.\ Lett.\ {\bf 97}, 186106 (2006)}.

\bibitem{Tangney_PRL04}
{P. Tangney, S.G. Louie, and M.L. Cohen, Phys.\ Rev.\ Lett.\ {\bf 93}, 065503
  (2004)}.

\bibitem{KC2005}
{ A.N. Kolmogorov and V.H. Crespi, Phys.\ Rev.\ B {\bf 71}, 235415 (2005)}.

\bibitem{Tangney_PRL06}
{P. Tangney, M.L. Cohen, and S.G. Louie, Phys.\ Rev.\ Lett.\ {\bf 97}, 195901
  (2006)}.

\bibitem{Berendsen}
{H.J.C. Berendsen, J.P.M. Postma, W.F. van Gunsteren, A. DiNola and J.R. Haak,
  J.\ Chem.\ Phys.\ {\bf 81} 3684, (1984)}.

\bibitem{Brenner}
{D.W. Brenner, Phys.\ Rev.\ B {\bf 42} 9458 (1990)}.

\bibitem{Andersen_JCP83}
{H.C. Andersen, J.\ Comput.\ Phys.\ {\bf 52} 24, (1983)}.

\bibitem{HUMP96}
William Humphrey, Andrew Dalke, and Klaus Schulten.
\newblock {VMD} -- {V}isual {M}olecular {D}ynamics.
\newblock {\em Journal of Molecular Graphics}, 14:33--38, 1996.

\end{thebibliography}


\newpage

\end{document}